\documentclass[journal=jacsat,manuscript=article]{achemso}

\usepackage{chemformula} 
\usepackage[T1]{fontenc} 

\author{Jason F. Khoury}
\affiliation{Department of Chemistry, Princeton University, Princeton, New Jersey 08544, United States}
\author{Bingzheng Han}
\affiliation{Department of Physics, Princeton University, Princeton, New Jersey 08544, United States}
\author{Milena Jovanovic}
\affiliation{Department of Chemistry, Princeton University, Princeton, New Jersey 08544, United States}
\author{Ratnadwip Singha}
\affiliation{Department of Chemistry, Princeton University, Princeton, New Jersey 08544, United States}
\author{Xiaoyu Song}
\affiliation{Department of Chemistry, Princeton University, Princeton, New Jersey 08544, United States}
\author{Raquel Queiroz}
\affiliation{Department of Physics, Columbia University, New York, New York 10027, United States}
\author{Nai-Phuan Ong}
\affiliation{Department of Physics, Princeton University, Princeton, New Jersey 08544, United States}
\author{Leslie M. Schoop}
\affiliation{Department of Chemistry, Princeton University, Princeton, New Jersey 08544, United States}
\email{lschoop@princeton.edu}
\phone{609-258-9390}

\title{A Class of Magnetic Topological Material Candidates with Hypervalent Bi Chains}

\begin{document}


\begin{abstract}
The link between crystal and electronic structure is crucial for understanding structure-property relations in solid-state chemistry. In particular, it has been instrumental in understanding topological materials, where electrons behave differently than they would in conventional solids. Herein, we identify 1D Bi chains as a structural motif of interest for topological materials. We focus on \ch{Sm3ZrBi5}, a new quasi-one-dimensional (1D) compound in the Ln$_3$MPn$_5$ (Ln = lanthanide; M = metal; Pn = pnictide) family that crystallizes in the \emph{P}$6_{3}$/\emph{mcm} space group. Density functional theory calculations indicate a complex, topologically non-trivial electronic structure that changes significantly in the presence of spin-orbit coupling. Magnetic measurements show a quasi-1D antiferromagnetic structure with two magnetic transitions at 11.7 and 10.7 K that are invariant to applied field up to 9 T, indicating magnetically frustrated spins. Heat capacity, electrical, and thermoelectric measurements support this claim and suggest complex scattering behavior in \ch{Sm3ZrBi5}. This work highlights 1D chains as an unexplored structural motif for identifying topological materials, as well as the potential for rich physical phenomena in the Ln$_3$MPn$_5$ family.
\end{abstract}

\section{Introduction}

Chemical bonding has proven to be important for the understanding and prediction of topological materials\cite{khoury2021chemical,schoop2018chemical, klemenz2019topological, klemenz2020role} --- a class of quantum matter that has been studied for several different reasons, such as high charge carrier mobility, high magnetoresistance, and symmetry-protected surface states.\cite{liu2014discovery, moll2016transport, liu2014stable, borisenko2014experimental, ali2014large, ali2015correlation, liang2015ultrahigh} Topological semimetals (TSMs) are a sub-class of topological matter that act as three-dimensional (3D) analogues of graphene, having linearly dispersed band crossings at the Fermi level.\cite{schoop2018tunable, liu2018giant, schnelle2013ferromagnetic} TSMs, unlike topological insulators (TIs), exhibit band crossings in the bulk band structure of a given material instead of as a surface state. If these states are gapped by spin-orbit coupling (SOC), a TI and topologically protected surface states could arise.

Due to significant fundamental and applied interest in topological materials, much work has been done to extensively catalogue and categorize available topological compounds, as well as predict potential candidates that have yet to be thoroughly investigated.\cite{vergniory2019complete, bradlyn2017topological, zhang2019catalogue} Such databases now serve as a guideline for researchers to synthesize new topological materials. While these databases are immensely useful, they also come with a few drawbacks: Firstly they do not provide any chemical heuristics to predict new topological materials. Secondly, they rely on generalized gradient approximation density functional theory (GGA-DFT), which by itself has known limitations about the accuracy of its band structure predictions.\cite{zunger2019beware, malyi2020realization} The limitations of such new predictions become evident when one considers that the most heavily studied topological materials still come from the same few structure types, such as square-nets, kagomes, and structures related to \ch{Bi2Te3}.\cite{liu2018giant, ye2018massive, liu2020orbital, kang2020dirac, chen2009experimental, ortiz2020cs, ortiz2021superconductivity} In addition, the databases revealed that the ``best'' topological materials had already been explored; tetradymite-type bismuth-chalcogenide TIs still have the largest band gaps and cleanest topological surface states.\cite{analytis2010bulk, analytis2010two, qu2010quantum, checkelsky2011bulk, hsieh2009tunable} Thus, the prediction of new, yet to be synthesized materials is desirable but can be very tricky, as the right combination of chemical intuition and computational knowledge is required to make accurate recommendations to the experimental community.\cite{zunger2019beware, malyi2020realization} Lastly, even if a candidate material \emph{is} topological and not yet investigated, it may be prohibitively difficult to synthesize in single crystalline form, as most experimental scientists gravitate toward crystals that are easy to grow and work with, allowing for physical characterization along multiple crystallographic directions. Therefore, it is crucial to provide direction for finding new materials by focusing on chemical features and heuristics in which previously overlooked topological candidates can be investigated.\cite{khoury2021chemical, klemenz2020role, khoury2019new, khoury2020subchalcogenides, gui2019new, isaeva2020crystal, hoffmann1987chemistry,  hirayama2018electrides, hirayama2020higher}

Hypervalent chemical bonds have been identified as one key feature for discovering new topological materials.\cite{klemenz2020role, khoury2021chemical} They are electron-rich, multicenter bonds where charge density is delocalized along a molecule or structural motif.\cite{klemenz2020role,khoury2021chemical, a2000hypervalent} In molecules, hypervalent bonding is a conventional three-center four-electron (3c--4e) interaction, such as \ch{I3-} or \ch{XeF2}.\cite{rundle1963problem, pimentel1951bonding} When charge balanced by another ion (typically an alkali or alkaline earth metal), this forms the basis for the Zintl-Klemm concept, where electron-precise, covalently bonded polyanions ionically interact with an electropositive cation.\cite{nesper1990structure, miller1996chemistry, toberer2010zintl, nesper2014zintl} In extended structural motifs, hypervalent bonding acts as an extension of the Zintl-Klemm concept, where an electron-rich motif (such as a hypervalent 1D chain or two-dimensional (2D) square-net) is charge balanced by other elements in the solid-state structure.\cite{a2000hypervalent} In square-nets, a hypervalent structural motif with two atoms per unit cell results in a folded band structure.\cite{klemenz2019topological, klemenz2020role, klemenz2020systematic} The band folding causes bands to cross and is the primary source of the topological bands.  A hypervalent bond will have two ``positive'' effects on a band structure. One, it stabilizes a half-filled band and thus will place the Fermi level at the band crossing or in a topological gap opened by SOC. Two, it comes along with greater orbital overlap as compared to metallic bonds, causing steep and widely dispersed bands, thus making the materials more robust to defects that could move the Fermi level. 

While an extensive amount of work has been done to showcase the potential of hypervalent 2D square-nets as an avenue for new topological materials,\cite{klemenz2020role, klemenz2019topological, klemenz2020systematic, weiland2019band, benavides2018casting} this idea has not yet been sufficiently extended to other lower dimensional, electron-rich structural motifs. Herein, we identify a new structural motif, hypervalent 1D Bi chains, as a building block for an unexplored class of magnetic topological materials. Here, too, each Bi chain consists of two atoms per unit cell, which induces band folding.
We identify a family of compounds, Ln$_3$MPn$_5$ (Ln = La--Nd; M = Ti, Zr, Hf, Mg, Mn, Nb; Pn = Bi, Sb), that hosts this structural motif.  We then identify one compound, \ch{Sm3ZrBi5}, to study in detail, showing that it is a new magnetic topological material with hypvervalent 1D Bi$^{2-}$ chains. \ch{Sm3ZrBi5} crystallizes in the \emph{P}$6_{3}$/\emph{mcm} space group with the \emph{anti}-\ch{Hf5Sn3Cu} structure type, and is a new member of the previously reported Ln$_3$MPn$_5$ family.\cite{ovchinnikov2018undistorted, murakami2017hypervalent, zelinska2008ternary, ferguson1997crystal, moore2002physical, bollore1995new} Density functional theory (DFT) calculations show a complex electronic structure with several band crossings at the Fermi level, some of which are gapped by SOC. Importantly, the features of a simple 1D chain model are found in the electronic structure. Magnetometry, heat capacity, electrical transport, and thermopower show two antiferromagnetic transitions at ~11.7 and 10.7 K that are invariant to changes in an applied magnetic field (in both measured field directions) up to 9 T. This unusual magnetic behavior appears to be the result of a complex quasi-1D magnetic structure, with frustrated spins that effectively ``pin'' both N\'eel temperatures in their respective positions. This work highlights the topological character of hypervalent 1D pnictide chain materials in the Ln$_3$MPn$_5$ family, as well as the interplay between complex magnetism and topology in  \ch{Sm3ZrBi5}.

\section{Experimental Section}

\subsection{Reagents.} Listed reagents were used as obtained: Bi pieces (99.999$\%$, Sigma-Aldrich), Zr sponge (>99$\%$, Sigma-Aldrich), and Sm pieces (99.9$\%$, Sigma-Aldrich).
\subsection{Synthesis.} \ch{Sm3ZrBi5} was synthesized via Bi flux reaction where Sm (3 mmol, 0.4511 g), Zr (1 mmol, 0.09120 g), and excess Bi (20 mmol, 4.120 g) were loaded into an alumina crucible and then placed into a 16 mm fused silica tube. A piece of silica wool was placed above the crucible in the tube to act as a filter for Bi flux upon centrifugation. The tube was flame sealed under vacuum with an Ar backfill at ~60 mTorr, heated to 1000 $^\circ$C over 10 h, held for 10 h, and cooled over the course of 72-96 h to 700 $^\circ$C. At 700 $^\circ$C, the tube was promptly removed from the furnace and centrifuged to separate the excess bismuth flux from the rod-like single crystals of \ch{Sm3ZrBi5}. Crystals grown via this method are on the order of several mm to ~1 cm in length, and will decompose in air if exposed for several hours. For transport, thermopower, and magnetic measurements, any residual Bi flux on the surface of the as-grown single crystals was scraped off with a blade. The elemental composition and approximate stoichiometry of the structure were confirmed via scanning electron microscopy (SEM)/energy-dispersive Spectroscopy (EDS) (Figure S1) and x-ray photoelectron spectroscopy (XPS) (Figure S2).
\subsection{Single Crystal X-ray Diffraction.} Single crystal diffraction data for \ch{Sm3ZrBi5} were collected on a Bruker D8 VENTURE with a PHOTON 3 CPAD detector using Mo K$\alpha$ radiation ($\lambda$ = 0.71073 {\AA}) with a graphite monochromator. Integrated data were corrected with a multiscan absorption correction using SADABS. The structures were solved via SHELXT using intrinsic phasing and refined with SHELXL using the least squares method.\cite{sheldrick2015shelxt} When the occupancy of Sm, Zr, and Bi Wyckoff positions were each freely refined, they were fully occupied; therefore, each site was fixed to be fully occupied. The crystallographic information can be found in Table S1 of the Supporting Information.
\subsection{Density Functional Theory (DFT) Calculations.} Band structure calculations on \ch{Sm3ZrBi5} were carried out using VASP 5.4.4. software. \cite{kresse1996efficiency, kresse1996efficient} The geometry was taken from the crystallographic data. Band structures were calculated using Perdew-Burke-Ernzerhof functional \cite{perdew1996generalized} for exchange and correlation. The calculations were done with and without spin-orbit coupling, using $\Gamma$-centered 15$\times$15$\times$15 Monkhorst-Pack mesh \cite{monkhorst1976special}, and Sm\_3, Zr\_sv, and Bi\_d Projector Augmented Wave (PAW) potentials. \cite{blochl1994projector, kresse1999ultrasoft} 
\subsection{Magnetometry.} Magnetic measurements were conducted using the vibrating sample magnetometer (VSM) option of either a Quantum Design DynaCool Physical Property Measurement System (PPMS) or a Magnetic Property Measurement System (MPMS) 3 SQUID magnetometer between 2 and 300 K. Single crystals were affixed to a quartz rod using GE Varnish. The field was applied parallel and perpendicular to the \emph{c} axis of \ch{Sm3ZrBi5}.
\subsection{Charge Transport.} Temperature-dependent resistivity ($\rho$) was measured on single crystals (with approximate dimensions of 3 x 0.3 x 0.3 $mm^3$) using a Quantum Design DynaCool PPMS from 2 to 300 K. Samples were measured by attaching Au wires with silver paste (Dupont 4929N) in a 4-point collinear geometry to a single crystal, with current along the \emph{c} direction and field perpendicular to the \emph{c} direction. A dilution refrigerator attachment was used to measure very low temperature transport down to 0.3 K.
\subsection{Heat Capacity.} Heat capacity ($C_p$) of \ch{Sm3ZrBi5} was measured using a Quantum Design DynaCool PPMS between 2 and 300 K, with Apiezon N grease attaching the samples to the stage. Variable-field heat capacity was conducted by applying a magnetic field perpendicular to the \emph{c} direction of \ch{Sm3ZrBi5}.
\subsection{Thermopower.} Thermopower (Seebeck coefficient) was measured with a home-built probe of twisted copper wires connecting to the contact pads of a cold finger to the copper clamps on the top of the probe. The vacuum of the probe is protected by an indium seal. The background temperature of the cold finger is measured and controlled with a Cernox 1050 thermometer attached to the heat sink, a heater on the cold finger, and a Lakeshore 340 Cryogenic Temperature Controller. The probe can be stabilized at temperature from around 4 K to room temperature in the PPMS. A 100 $\Omega$ heater is attached to the top of the crystal with silver paint. Current is later applied to the heater with a Keithley 6221 current source to introduce a temperature gradient (\emph{$-\nabla$T}) throughout the crystal. The temperature difference (\emph{$\Delta$T}) is measured with a Lakeshore 340 Cryogenic Temperature Controller and two Cernox 1050 thermometers attached to the top and bottom of the crystal, respectively. The thermopower signal contacts are made with 25 $\mu$m phosphor bronze (ph-bronze) wires. Ph-bronze wires are poor conductors of heat because of their negligible Seebeck coefficient, thus reducing the heat leakage between the sample and the cold finger through the wires. The Ph-bronze wires are attached to the contact pads on the cold finger of the probe, which are connected all the way up to the top of the probe through continuous twisted copper wires, and the thermopower signal (V) is measured with a Keithley 2182 nanovoltmeter. The Seebeck coefficient is the given by $S=\frac{V}{\Delta T}$.

\section{Results and Discussion}
\subsection{Synthesis and Structure.} 

\begin{figure}[hbt!]
  \includegraphics[width=\textwidth]{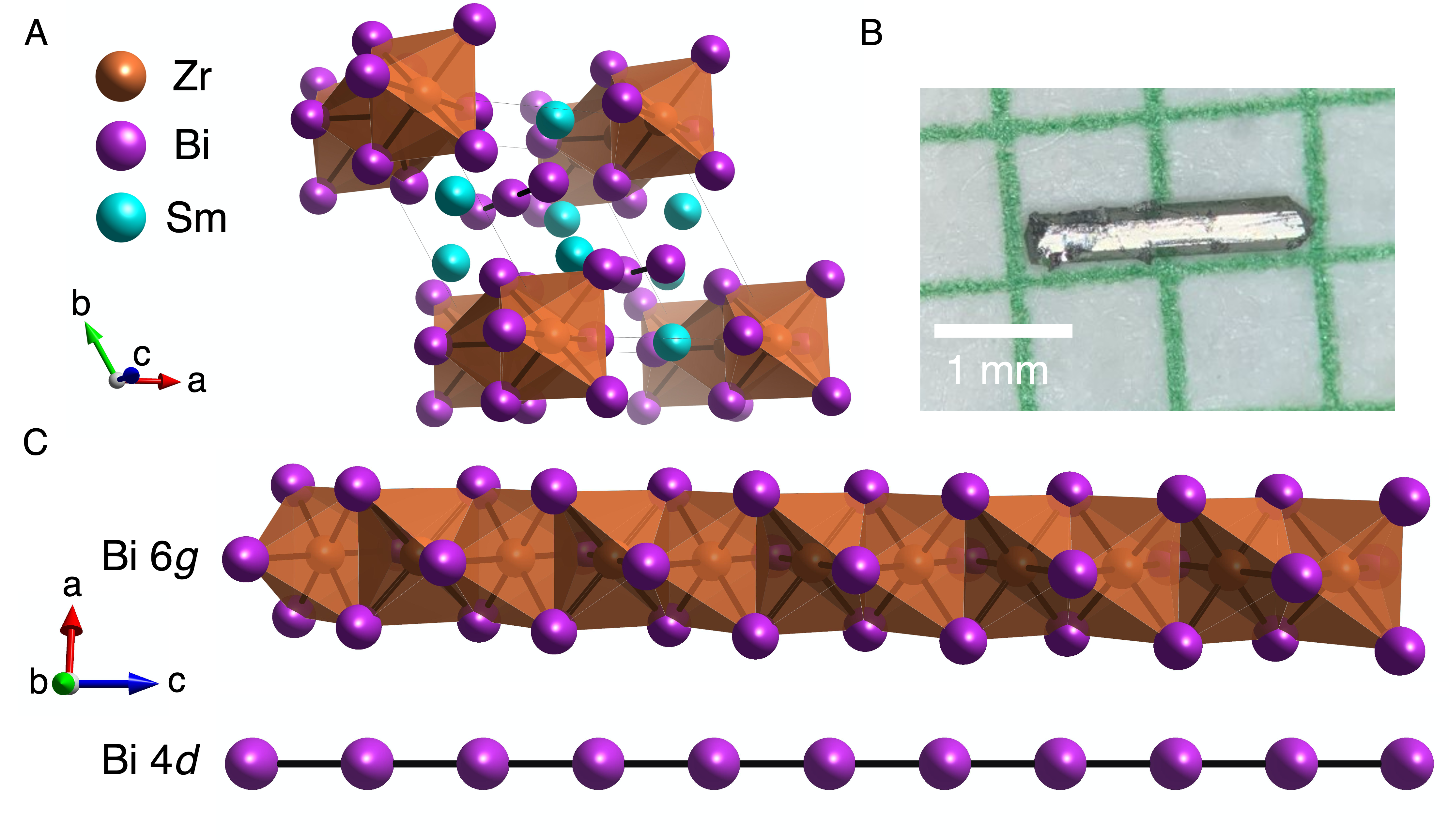}
  \caption{Structure of \ch{Sm3ZrBi5}. (A) Individual unit cell of \ch{Sm3ZrBi5} along the \emph{c} axis. (B) Flux-grown crystal of \ch{Sm3ZrBi5}, highlighting the hexagonal rod-like geometry. (C) Side view of the separate 1D \ch{ZrBi6} face-sharing octahedral (top) and the Bi$^{2-}$ chain (bottom) motifs. The Wyckoff position of each Bi atom is shown next to the motif.}
  \label{Structure}
  \centering
\end{figure}

\ch{Sm3ZrBi5} is a new compound with a quasi-1D structure, and it crystallizes in the \emph{P}$6_{3}$/\emph{mcm} space group hosting the \emph{anti}-\ch{Hf5Sn3Cu} structure type, as seen in Figure \ref{Structure}. Crystallographic information can be found in Table S1. The structure is part of the Ln$_3$MPn$_5$ family (Ln = La--Nd; M = Ti, Zr, Hf, Mg, Mn, Nb; Pn = Bi, Sb), which has rich chemical flexibility, resulting in many complex magnetic phases.\cite{matin2017probing, ritter2021magnetic, motoyama2018magnetic, shinozaki2020study} \ch{Sm3ZrBi5} hosts 1D Bi chains along the \emph{c} axis, with Bi--Bi bond lengths of 3.20790(16) {\AA}, much longer than those seen in elemental Bi (3.0712(11) {\AA}). The bonding of the chains can be understood as hypervalent, as previously discussed by Murakami and coworkers, where Bi$^{2-}$ anions have seven electrons per atom, causing electron-rich multicenter bonding.\cite{murakami2017hypervalent} These longer Bi--Bi bonds are attributable to their hypervalent nature, as conventional two-center two-electron (2c--2e) bonds tend to have closer, more localized interactions.\cite{khoury2021chemical, a2000hypervalent, hoffmann1987chemistry} A unit cell contains two chains, and each chain will bring two atoms to the unit cell, which makes for a total of four Bi chain atoms per unit cell.

\begin{figure}[hbt!]
  \includegraphics[width=\textwidth]{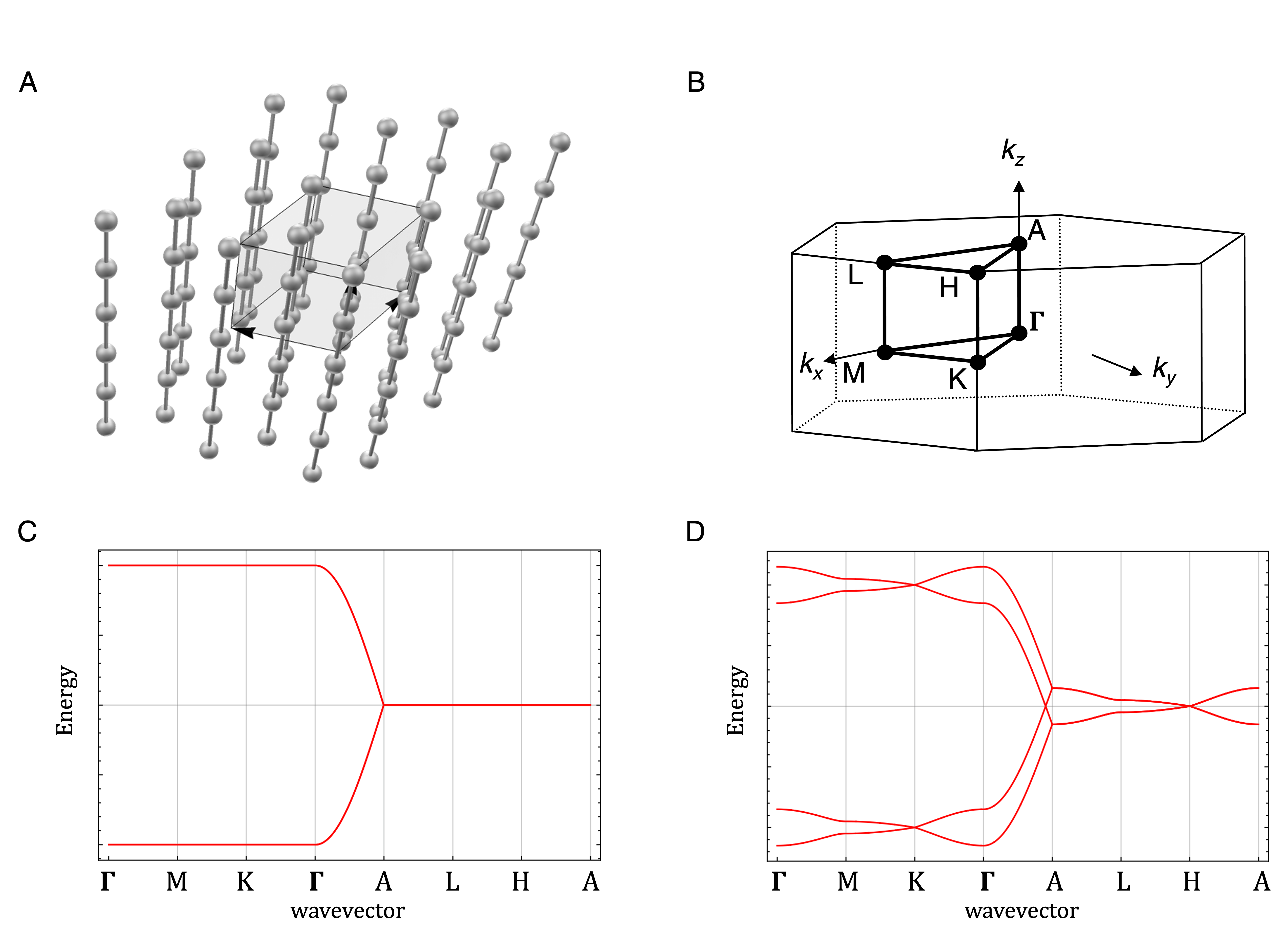}
  \caption{Scheme of hypervalent chains and the electronic structure of a half-filled $p_z$ band without SOC. (A) Unit cell of 1D chains, as they appear in the \ch{Sm3ZrBi5} structure. (B) Hexagonal Brillouin zone path. (C) Electronic structure with non-interacting chains, showing very dispersive bands along $\Gamma$--A ($k_z$) and flat bands along L--H--A ($k_z = \pi$ plane). (D) Electronic structure with interacting chains, showing the band crossing along $\Gamma$--A and the non-symmorphic degeneracies at the A point. }
  \label{Scheme}
  \centering
\end{figure}

Hypervalent 1D chains such as Bi$^{2-}$ have been suggested to express topological character, as a half-filled band of $p_z$ orbitals allows for the stabilization of symmetry-protected band crossings in a solid-state structure.\cite{schoop2018chemical, khoury2021chemical} One would normally expect such a structural pattern with half-filled bands to distort, and the origin of these distortions have been studied in other 1D compounds.\cite{a2000hypervalent, hoffmann1987chemistry, shi2021charge} To demonstrate this idea, we show a simple model of linear chains with a half-filled $p_z$ band, and two atoms per unit cell in Figure \ref{Scheme}. For this model, we used the \ch{Sm3ZrBi5} structure as a basis but removed all atoms except the linear Bi chains (Figure \ref{Scheme}A--B). The model contains only $p_z$ orbitals, but the chains are assumed to not interact with each other in Figure \ref{Scheme}C. A 1D chain of $p_z$ orbitals offers an intuitive model for inorganic chemists to follow, showing that the link between crystal and electronic structure is attributable to simple structural patterns in the unit cell. In \ref{Scheme}D, interactions between the two chains are allowed. The non-interacting chain band structure exhibits very dispersive bands along $\Gamma$--A ($k_z$, inverse \emph{c} axis), a degeneracy at A as a result of band folding, and flat bands along L--H--A ($k_z = \pi$ plane). Thus, linear chains with half-filled $p_z$ orbitals as they appear in \ch{Sm3ZrBi5} can lead to not only dispersive linear bands, but also flat bands, the latter of which has often been associated with electron-electron correlations in materials.\cite{regnault2021catalogue} Chemical intuition often links flat bands with non-bonding orbitals or lone pairs. In this case, however, the flat band is a direct consequence of the hypervalent 1D chains, as shown in our tight-binding model in Figure \ref{Scheme}C--D. They stem from different interactions in separate regions of the Brillouin zone, such as the $k_z = 0$ and $k_z = \pi$ planes. This band structure, while exhibiting disperse bands that cross at A, is not topological. However, when the chains interact (Figure \ref{Scheme}D), a topological band crossing is induced along the $\Gamma$--A line. This serves as a basis to understand the topological character of \ch{Sm3ZrBi5}. Similarly, in the DFT model shown below, the two chains also interact, inducing topological bands.

Nonetheless, \ch{Sm3ZrBi5} contains more than just hypervalent Bi chains. In addition, the structure contains face-sharing \ch{ZrBi6} octahedra form 1D chains that also run along the \emph{c} axis, with Zr-Bi bond lengths of 3.0156(9) {\AA}. The Bi-Bi distances in the face-sharing octahedra are $\sim$4.1--4.4 {\AA}, resulting in weaker but still significant bonding interactions. The Sm atoms are arranged in a distorted 3D hexagonal motif, and do not appear to have any close bonding interactions with either the Zr or Bi atoms in the structure, acting primarily as an ionic, charge-balancing sublattice. 

In the Ln$_3$MPn$_5$ family, we can count electrons and charges due to the chemical bonding of Pn and the ionic nature of the Ln lattice. This allows for a greater understanding of the bonding, electronic structure, and physical properties of \ch{Sm3ZrBi5}. The Sm cations are expected to adopt the 3+ oxidation state, such as most other lanthanides. Bi has two unique Wyckoff positions in the structure, which can be assigned separate anionic oxidation states: on the chain (2--, 4$d$) and the octahedra (6$g$). One would expect the latter to adopt a formal oxidation state of 3--, which is typical of bismuthide Zintl-Klemm phases.\cite{miller1996chemistry, toberer2010zintl} In this case, the Zr cation would have to be 4+ to maintain charge balance.\cite{teufer1962crystal} However, there is compelling evidence that Zr may be in the 2+ state. First, the Ln$_3$MPn$_5$ structure can accommodate octahedrally coordinated Mg$^{2+}$ in the \emph{M} site, with the other oxidation states behaving similarly, as seen in \ch{La3MgBi5}.\cite{pan2006synthesis} The relative stability of a 2+ cation in the structure suggests that the 6$g$ Bi ligand atoms may have only partially filled $p$ orbitals. The flexibility of assigned oxidation states in the \emph{M} site of Ln$_3$MPn$_5$ is indicative of a structure that does not fully conform to the Zintl-Klemm concept, which is attributable to either delocalized bonding in the \ch{ZrBi6} octahedra or mixed valency on the metal site. XPS data for Zr (Figure S2) suggests a single oxidation state, ruling out mixed valency in \ch{Sm3ZrBi5}, but does not provide conclusive evidence of 2+ or 4+, as the ligand environment of the \ch{ZrBi6} octahedra causes a considerable shift of the binding energy relative to known standard materials. The final reason that the Zr oxidation state may not be 4+ is that DFT projected bands of the \ch{Sm3ZrBi5} structure (Figure S3--4) show partially occupied Bi states from the 6$g$ Bi anions at the Fermi level, which would not occur if the electron count was exact, as seen in traditional bismuthides.\cite{derrien2002synthesis, murakami2017hypervalent, zelinska2008ternary} As we show below, the partially vacant 6$g$ Bi orbitals have important consequences for the topological character of the band structure.

\subsection{Electronic Structure.}

\begin{figure}[hbt!]
  \includegraphics[width=\textwidth]{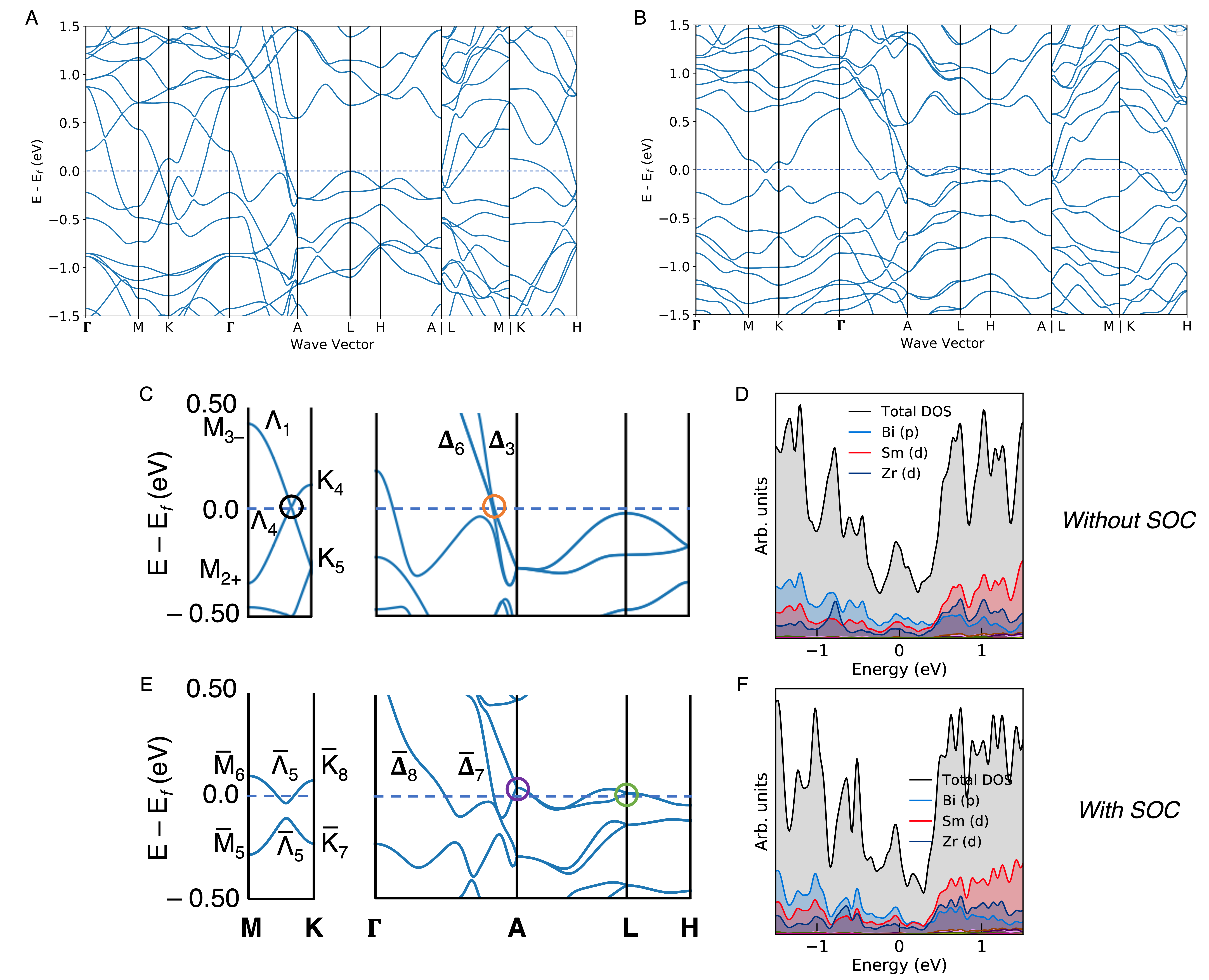}
  \caption{Electronic band structure of \ch{Sm3ZrBi5} without (A, C) and with (B, E) spin-orbit coupling (SOC). Enlarged band structures without (C) and with (E) SOC and their irreducible representations (irreps) show the topologically protected crossings (black and orange circles in panel C) and their transformations. Enforced non-symmorphic degeneracies (purple and green circles) near the Fermi level are circled in panel D. Total and partial density of states (DOS) calculations without (D) and with (F) SOC, highlighting the large contribution of Bi $p$ states at the Fermi level. The overall topological character of the material is a result of several complex Bi--Bi interactions parallel and perpendicular to the \emph{c} direction.}
  \label{BandStructure}
  \centering
\end{figure}

The electronic band structure of \ch{Sm3ZrBi5}, calculated with and without SOC, is shown in Figure \ref{BandStructure}. Projections of Bi orbitals on the DFT band structure, resolved by Wyckoff positions (Bi$^{2-}$ chains, 4$d$, versus Bi anions from the \ch{ZrBi6} octahedra, 6$g$), can be found in Figure S3--4 of the Supporting Information. Bands resembling the 1D-chain model can be recognized in the steep dispersion along the $\Gamma$--A line and flat bands in the $k_z = \pi$ plane (along A--L--H--A). Figures S3--4 show that these bands indeed have significant contribution from the 4$d$ Bi$^{2-}$ chain atoms.

The band structure without SOC (Figure \ref{BandStructure}A, C--D) exhibits topologically protected band crossings near the Fermi level along the $\Gamma$--A (orange circle in Figure \ref{BandStructure}C) and M--K (black circle) directions. The $\Gamma$--A line preserves a 6$mm$ ($C_{6v}$) point group symmetry, and the mirror plane protects a crossing of non-degenerate $\Delta_3$ ($B_2$) and doubly-degenerate $\Delta_6$ ($E_2$) bands, located at the Fermi level.\cite{server2018check} Bi $p_z$ orbitals on the 4$d$ Wyckoff position contribute significantly to $\Delta_3$, which is expected given that the 1D chains run along the \emph{c} axis (Figure S3--4). $\Delta_6$, on the other hand, has orbital contributions from both the 6$g$ and 4$d$ Bi atoms, suggesting more complex interactions in the overall unit cell are involved. The shape and position of the steep bands along $\Gamma$--A generally agree with the 1D chain model schematically shown in Figure \ref{Scheme}D. However, the overall topological character of the electronic structure requires taking into account contributions from Bi atoms in the \ch{ZrBi6} octahedra. 

Another symmetry-protected crossing near the Fermi level is visible along the M--K direction. The bands in this region are mainly composed of 6$g$ Bi's $p_x$ and $p_y$ states (\emph{i.e} from the face-sharing \ch{ZrBi6} octahedra).  
The M--K line has in the $mm$2 ($C_{2v}$) point group symmetry, and mixing between bands of $\Lambda_1$ ($A_1$) and $\Lambda_4$ ($B_2$) symmetries is forbidden. The bands surrounding the crossing along M--K are very dispersive, with a band width of approximately 0.8 eV, allowing for robust topological behavior that is invariant to defects that would alter the position of the Fermi level.\cite{schoop2016dirac} 

Furthermore, non-symmorphic symmetry elements in the \emph{P}$6_{3}$/\emph{mcm} space group enforce four-fold degeneracy of bands along the H--A and A--L high symmetry lines. These non-symmorphic symmetry elements are a glide plane that interchanges $k_x$ and $k_y$, and another one which reverses the sign of $k_y$. Both are followed by a translation in $z$, that gives a complex phase of $i$ to its eigenvalues at $k_z = \pi$. These enforced degeneracies can be seen, for example, at the A point $\sim$0.3 eV below the Fermi level. The total and partial density of states (DOS) without SOC (Figure \ref{BandStructure}D) shows a small, broad peak at the Fermi level, primarily from Bi $p$ orbitals, with smaller contributions from Zr $d$ and Sm $d$ states. This is consistent with the projected band structures in Figure S3--4, showing that there are significant Bi $p$ contributions from bands that cross the Fermi level. Taken together, the electronic structure without SOC shows that the bands from the hypervalent linear chain model, shown in Figure \ref{Scheme}, appear in the DFT band structure of a real material with such chains, i.e \ch{Sm3ZrBi5}. In addition, there are also Bi--Bi interactions within the \emph{ab} plane that are important to the overall band structure and its topological character. However, all topological bands are derived from Bi orbitals, suggesting that SOC will have a dramatic effect on the band structure.

When SOC is accounted for, the band structure changes significantly (Figure \ref{BandStructure}B, E--F). These changes include a local gap formation along the M--K band crossing (Figure \ref{BandStructure}C, E), a shift of the crossing along $\Gamma$--A away from the Fermi level, and a shift of non-symmorphic symmetry-enforced degeneracies at the A and L points toward the Fermi level, which notably ``closes'' the band gap that was previously in the $k_z = \pi$ plane. SOC requires double group symmetry, which differ from conventional point groups in that they account for electron spin, causing the irreducible representations (irreps) of the bands to change. This can be seen, for example along the M--K line (within the $k_z = 0$ plane). The crossing, which was previously between bands with the $\Lambda_1$ and $\Lambda_4$ irreps, is gapped by SOC because both bands are now described by the $\bar\Lambda_5$ irrep. The resulting topologically protected local energy gap ($E_g$) along the M--K direction is $\sim$0.06 eV, which is smaller than those of most known topological insulators, but still significant. In addition, the other directions in the $k_z = 0$ plane (\emph{i.e.} $\Gamma$--M and $\Gamma$--K) had several bands crossing the Fermi level that are now gapped by SOC. 

The band crossings along the $\Gamma$--A line, which were previously between bands with the $\Delta_3$ and $\Delta_6$ irreps (Figure \ref{BandStructure}C), are rearranged in the SOC band structure (Figure \ref{BandStructure}E) since combined with the spin representation, the two bands now transform under the spinful representations $\Delta_3\to \bar \Delta_7$ and $\Delta_6\to \bar\Delta_7+\bar\Delta_8$. As a consequence, the bands move and intermix significantly along the $\Gamma$--A line such that two new irreps, $\bar\Delta_7$ and $\bar\Delta_8$, replace the original bands near the Fermi level. Other bands along the $k_z$ direction, \emph{i.e} the L--M and K--H symmetry lines, also undergo changes, as there are no longer any crossings along L--M and some of the bands shift to create a new crossing near the Fermi level along K--H. SOC also introduces major changes in the $k_z = \pi$ plane (\emph{i.e.} A--L, L--H, and H--A), shifting multiple bands with non-symmorphically protected degeneracies to the Fermi level. The enforced degeneracies are at the A (purple circle in Figure \ref{BandStructure}E) and L (green circle) points. Notably, SOC significantly flattens these bands, leading to a higher density of states at the Fermi level, potentially allowing for the interplay of electron correlations and topology. The total and partial DOS with SOC (Figure \ref{BandStructure}F) show a sharper peak than in Figure \ref{BandStructure}D (without SOC), but shifted slightly below the Fermi level. The sharpness of the peak is likely attributable to the flattening of the bands in the $k_z = \pi$ plane. The relative orbital contributions are also similar to Figure \ref{BandStructure}D and Figure S3--4, with Bi $p$ states being the largest contributor, along with smaller contributions from Zr $d$ and Sm $d$ orbitals.

\subsection{Magnetism.}
\begin{figure}[hbt!]
  \includegraphics[width=\textwidth]{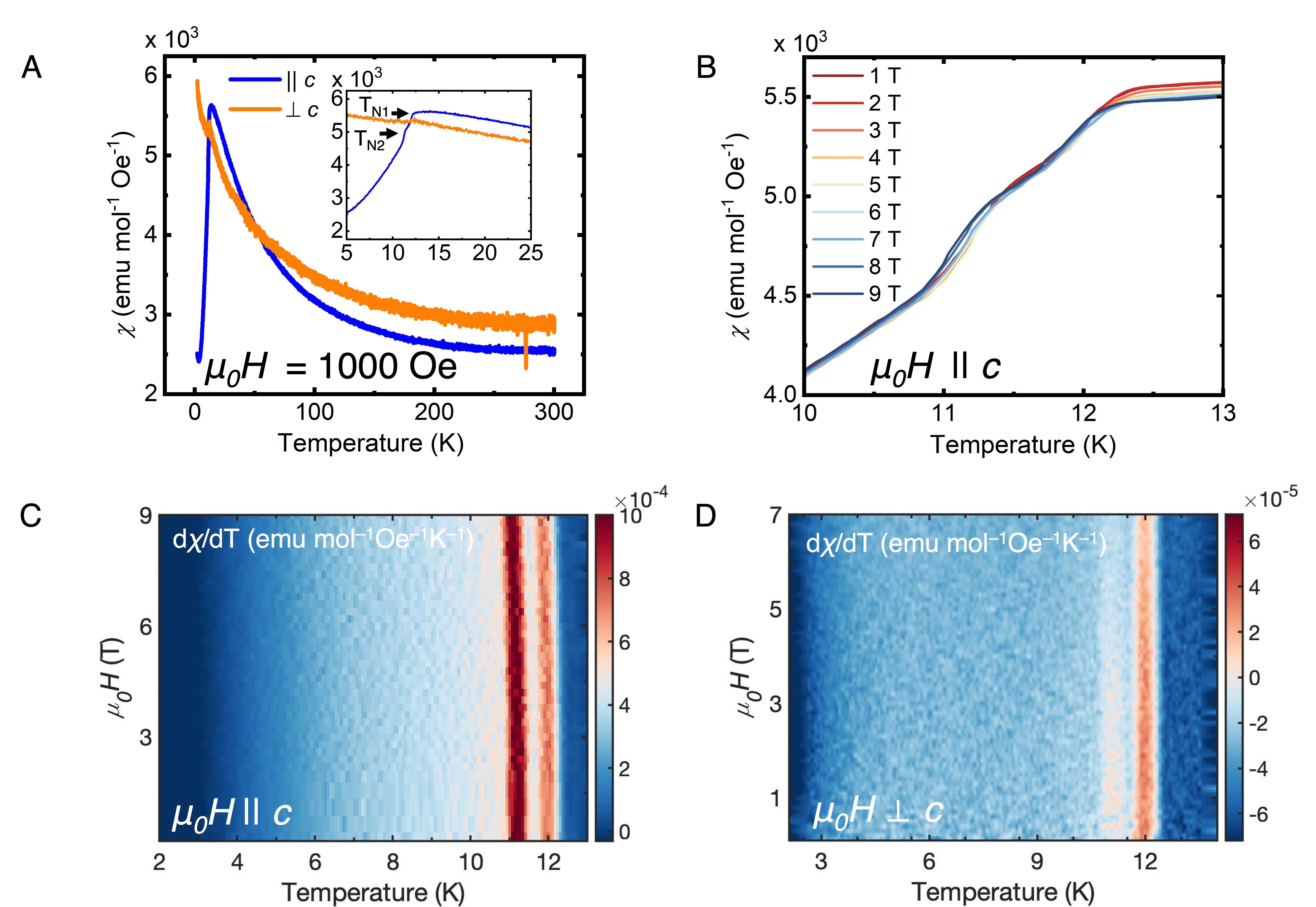}
  \caption{Magnetic properties of \ch{Sm3ZrBi5} with field applied parallel and perpendicular to the \emph{c} direction. (A) Magnetic susceptibility (\emph{$\chi$}) with field applied parallel and perpendicular to the \emph{c} direction. (B) Magnetic susceptibility as a function of varying fields near the two magnetic transitions with field along the \emph{c} direction. The transition temperatures are very constant with field applied along both directions, which can be seen in the magnetic phase diagrams as a function of the derivative of magnetic susceptibility (d\emph{$\chi$}/dT) with respect to temperature (C and D).}
  \label{Magnetism}
  \centering
\end{figure}

Now that we have confirmed the topological character of \ch{Sm3ZrBi5}, driven by its Bi--Bi interactions, it is of interest to study its magnetic properties, which are expected to arise from the presence of Sm. The interplay of topology and magnetism is of widespread interest, as it can lead to Fermi arc surface states in Weyl semimetals (WSMs) along with transport phenomena such as the chiral anomaly and large intrinsic anomalous Hall effects.\cite{liu2018giant, zyuzin2012topological, schoop2018tunable} To assess the magnetic properties, we measured the  temperature and field-dependent magnetic susceptibility, as displayed in Figure \ref{Magnetism}. Figure \ref{Magnetism}A shows the magnetic susceptibility (\emph{$\chi$}) with respect to temperature. Two broad antiferromagnetic (AFM) transitions are visible if the field is applied along the \emph{c} axis, with N\'eel temperatures ($T_N$) at ~12.2 and ~11.4 K. The \emph{c} axis resembles the easy axis of magnetization, as the same AFM transitions are still present but much less noticeable along the \emph{ab} plane. The anisotropic magnetic structure is in good agreement with the 1D nature of the crystal structure. Curie-Weiss fits to the data are obtained following the expression $\emph{$\chi$} = \frac{C}{T - \emph{$\theta_{CW}$}}$, where \emph{$\chi$} is magnetic susceptibility, C is the Curie constant, T is the absolute temperature, and \emph{$\theta_{CW}$} is the Curie temperature. A fitting range between approximately 15 and 45 K was applied, as Sm exhibits large, temperature-independent Van Vleck paramagnetism from the first excited multiplet state (J = 7/2) from Sm$^{3+}$ cations, which requires a low temperature region for the Curie-Weiss fit.\cite{malik1974crystal} The fitting parameters along both field directions (Figure S5) result in negative Curie temperatures (\emph{$\theta_{CW}$} = --76.82 and --116.90 K when the field is applied parallel and perpendicular to the \emph{c} axis, respectively), suggesting AFM interactions. The fitted magnetic moments (\emph{$\mu_{eff}$}) are 1.18 and 1.34 \emph{$\mu_B$}/mol$_{Sm}$ when the field is applied parallel and perpendicular to the \emph{c} axis, respectively. These moments are higher than the predicted free-ion value of 0.85 \emph{$\mu_B$} for a Sm$^{3+}$ cation or those observed in some Sm compounds, such as Sm-pyrochlores.\cite{sanders2016re, singh2008manifestation}

Figure \ref{Magnetism}B shows the temperature-dependent magnetic susceptibility (with field parallel to \emph{c}) with multiple field strengths up to 9 T. Surprisingly, the two N\'eel temperatures are almost constant as a function of applied field; most AFM transitions shift to lower temperatures when fields are applied, which stabilizes the AFM phase. This invariance to field strength is also reflected in the magnetic phase diagrams of \ch{Sm3ZrBi5} (Figure \ref{Magnetism}C--D), where the d\emph{$\chi$}/dT data show that both transitions are effectively constant as a function of magnetic field along both measured directions. Additional magnetic phase diagrams are shown in Figure S5 of the Supporting Information. 

One possible reason for why the magnetic structure of \ch{Sm3ZrBi5} is both anisotropic yet invariant to magnetic field (up to 9 T) is due to a complex, quasi-1D magnetic structure, where the spins are strongly coupled due to magnetic frustration. A system with similar magnetic behavior is the $S=\frac{1}{2}$ antiferromagnet \ch{Li2CuW2O8}. Here, the N\'eel transition is also invariant to applied field up to 9 T in polycrystalline samples, which is believed to be caused by a magnetically frustrated 3D structure that mimics quasi-1D behavior across a certain temperature range.\cite{ranjith2015collinear} In the case of \ch{Li2CuW2O8}, the magnetic frustration may result in a high saturation field ($\mu_S$) of 29 T.\cite{ranjith2015collinear} It is possible that a very high applied field might be necessary to saturate the magnetization of \ch{Sm3ZrBi5}. However, \ch{Sm3ZrBi5} is distinct from \ch{Li2CuW2O8} in several ways. Firstly, it has two $T_N$ transitions as opposed to one, suggesting a more complex magnetic ground state. Secondly, the constant N\'eel temperature is observed when the field is applied along multiple directions, which was not observable in the measurements of their polycrystalline samples. Due to the high cross-sectional absorption of Sm$^{3+}$ for neutron diffraction, it is difficult to measure an accurate magnetic structure solution, but high-field measurements and with Muon spin resonance spectroscopy ($\mu$SR) would be a promising way to further understand the complex magnetic behavior of this system.

\subsection{Heat Capacity.}

\begin{figure}[hbt!]
  \includegraphics[width=\textwidth]{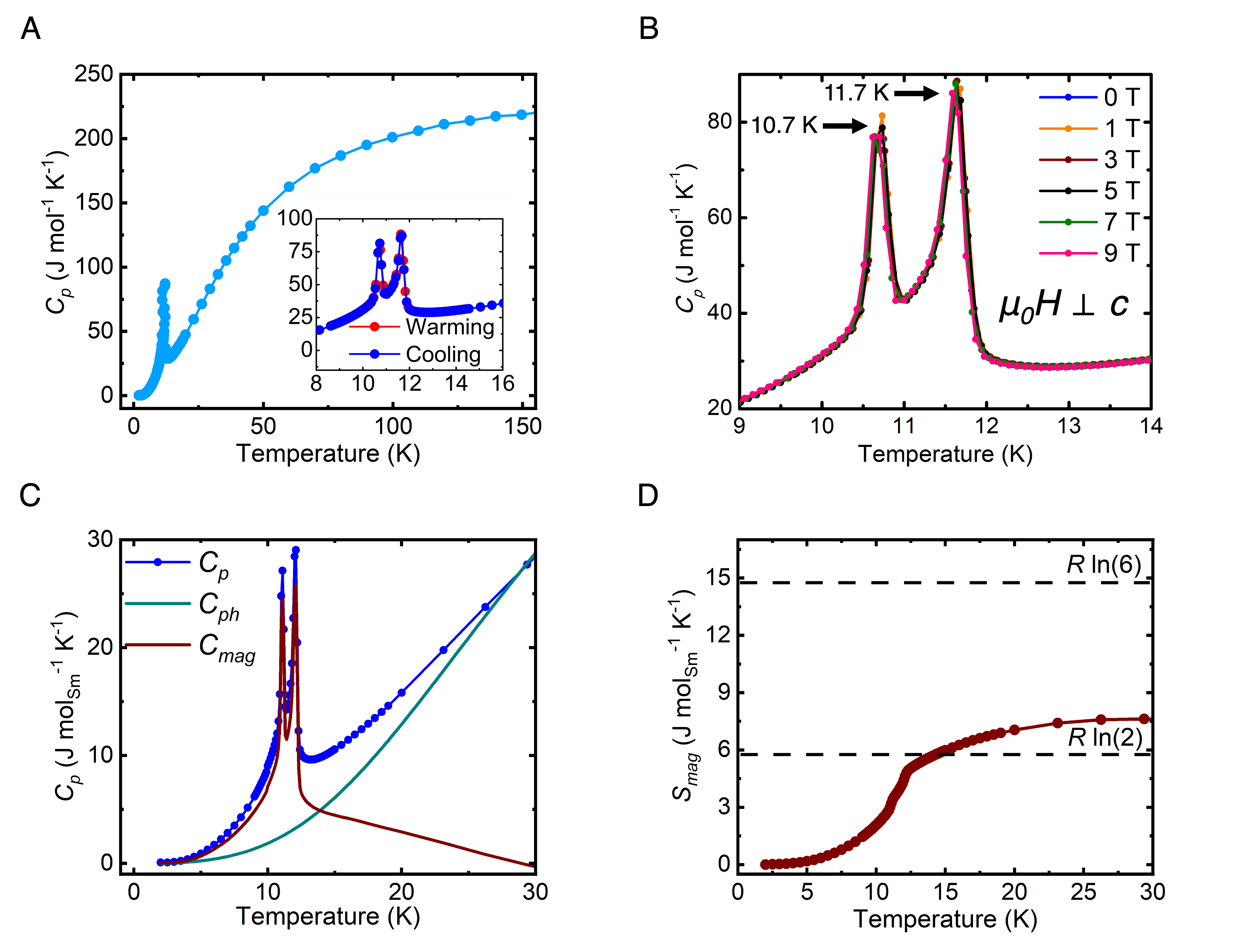}
  \caption{Heat capacity ($C_p$) of \ch{Sm3ZrBi5}. (A) Temperature-dependent Heat capacity, showing the two AFM transitions at ~11.7 and 10.7 K. The inset shows that there is no thermal hysteresis from the transitions, supporting that they are magnetic in nature. (B) $C_p$ under multiple magnetic fields (up to 9 T) from 9 to 14 K. The transitions have negligible changes as a function of high applied field, suggesting a complex, strongly coupled magnetic structure. (C) Debye fit for the phonon heat capacity ($C_{ph}$), which can be used to derive the magnetic heat capacity ($C_{mag}$) by subtracting the phonon background from total $C_p$. (D) Magnetic entropy ($S_{mag}$) integrated from $\frac{C_{mag}}{T}$. The magnetic entropy saturates close to R$\ln$2, suggesting a $S=\frac{1}{2}$ state as opposed to a $S=\frac{5}{2}$ (R$\ln$6) state in a free Sm$^{3+}$ ion.}
  \label{cp}
  \centering
\end{figure}

To continue to investigate the nature of these magnetic transitions and their invariance to applied field, we performed heat capacity($C_p$) measurements to precisely identify the temperature of each transition and their magnetic entropy. Heat capacity data for \ch{Sm3ZrBi5} can be viewed in Figure \ref{cp}.  Two distinct transitions appear at approximately 11.7 and 10.7 K (Figure \ref{cp}A). These transitions show no thermal hysteresis, as shown in the inset of Figure \ref{cp}A, supporting that the peaks are magnetic rather than structural. Similarly, as evidenced in the magnetic susceptibility, the peaks in the heat capacity do not shift with applied field up to 9 T (Figure \ref{cp}B). From the heat capacity data, the spin state of \ch{Sm3ZrBi5} can be determined by integrating the magnetic contribution of the total heat capacity over temperature ($\frac{C_{mag}}{T}$) to calculate the total magnetic entropy ($S_{mag}$). To determine the value of $C_{mag}$, the phonon contribution to the heat capacity ($C_{ph}$) must be subtracted from the total $C_p$. We can fit the value of $C_{ph}$ using the Debye equation shown below:

\[C_{ph} = 9R(\frac{T}{\theta_D})^3\int\limits_0^\frac{\theta_D}{T} \frac{x^4e^x}{(e^x-1)^2} \ dx \]

where R is the ideal gas constant, T is temperature, and $\theta_D$ is the Debye temperature. $C_{ph}$ and $C_{mag}$ are shown in Figure \ref{cp}C, and the fitted Debye temperature of \ch{Sm3ZrBi5} is approximately 147 K. By integrating $\frac{C_{mag}}{T}$ with respect to temperature, the total $S_{mag}$ (Figure \ref{cp}D) saturates at around 7.57 Jmol$^{-1}$K$^{-1}$, much closer to $S=\frac{1}{2}$ (R$\ln$2) as opposed to the $S=\frac{5}{2}$ (R$\ln$6) spin state expected for a free Sm$^{3+}$ ion. There is precedent for intermetallics containing Sm$^{3+}$ to be in a $S=\frac{1}{2}$ doublet state due to crystal field effects, with examples of \ch{SmPd2Al3} and \ch{SmPtSi2}.\cite{pospivsil2010samarium, yamaguchi2018fundamental}. The value of the $S_{mag}$ is higher than R$\ln$2 (5.76 Jmol$^{-1}$K$^{-1}$) may be attributable to fitting the complex phonon backgrounds of rare earth intermetallics, since similar fitting issues have been reported for \ch{SmPd2Al3}.\cite{pospivsil2010samarium} It is also possible that the magnetic ground state is $S=1$ because it is reasonably close to the value of R$\ln$3 (9.13 Jmol$^{-1}$K$^{-1}$), but to the best of our knowledge, this spin state has not been reported for Sm intermetallics.

\subsection{Charge Transport.}
\begin{figure}[hbt!]
  \includegraphics[width=0.45\textwidth]{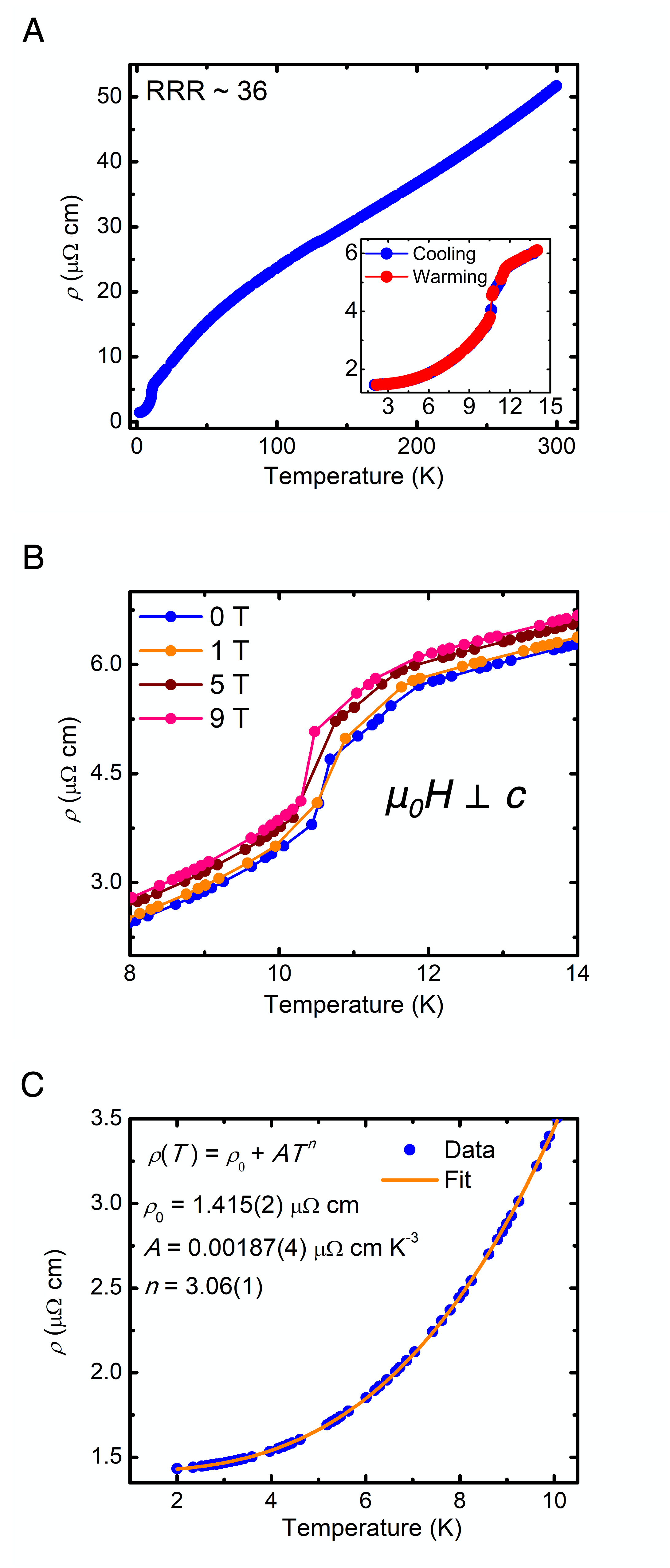}
  \caption{Resistivity ($\rho$) of \ch{Sm3ZrBi5}. (A) Temperature-dependent resistivity with current applied along the \emph{c} axis. The low temperature inset shows two magnetic transitions at 11.7 and 10.7 K. There is no thermal hysteresis from the transitions, which is typical for reversible magnetic ordering. (B) Variable field resistivity data (up to 9 T) from 8 to 14 K, showing that there is negligible change in the electron transport, corroborating the magnetic susceptibility and heat capacity data. (C) Fitted low temperature resistivity, revealing a power law of 3, suggesting electron--magnon or interband electron--phonon scattering.}
  \label{Transport}
  \centering
\end{figure}

Now that we have thoroughly characterized the nature of the magnetic transitions in \ch{Sm3ZrBi5}, we can use electrical transport measurements to corroborate their invariance to applied field. Transport measurements also give insight into the nature of the electron scattering mechanism of the compound, since topological materials tend to have unconventional scattering behavior.\cite{tafti2016resistivity, sun2016large, singha2017large} The temperature-dependent electrical resistivity, measured with the current applied along the \emph{c} axis, is shown in Figure \ref{Transport}A. \ch{Sm3ZrBi5} expresses metallic behavior, since the resistivity decreases monotonically with decreasing temperature. The residual resistivity ratio (RRR = $\rho_{300K}$/$\rho_{2K}$) of the crystal is approximately 36, indicating good crystal quality.  The inset of Figure \ref{Transport}A shows two AFM transitions at 11.7 and 10.7 K, which are consistent with the heat capacity data from Figure \ref{cp}. There is no thermal hysteresis in the transitions, further supporting that they are magnetic rather than structural. Figure \ref{Transport}B displays the low temperature resistivity with field applied perpendicular to the \emph{c} axis up to 9 T, revealing almost negligible change in both magnetic transitions, in agreement with the magnetic and heat capacity data. Dilution refrigerator measurements down to 0.3 K show no additional magnetic or superconducting transitions in \ch{Sm3ZrBi5}, shown in Figure S6. Low temperature resistivity data from 2 to 10 K (see Figure \ref{Transport}C) was fitted to the equation $\emph{$\rho$(T)}$ = $\rho_0$ + $AT^n$, where $\rho_0$ is the residual resistivity, $A$ is the scaling factor, $T$ is temperature, and $n$ is the power law for the scattering mechanism. The value of $n$ is almost exactly 3, revealing a $T^3$-dependence of the low temperature resistivity, suggesting electron--magnon (due to magnetic ordering) or interband s--d electron--phonon scattering as the primary scattering mechanism.\cite{singha2017large, sun2016large} This power law is distinct from quadratic ($T^2$) behavior for a typical Fermi liquid dominated by electron--electron scattering or $T^5$ behavior where conventional electron--phonon scattering is the dominant mechanism.\cite{calta2014four}

\subsection{Thermopower.}

\begin{figure}[hbt!]
  \includegraphics[width=0.45\textwidth]{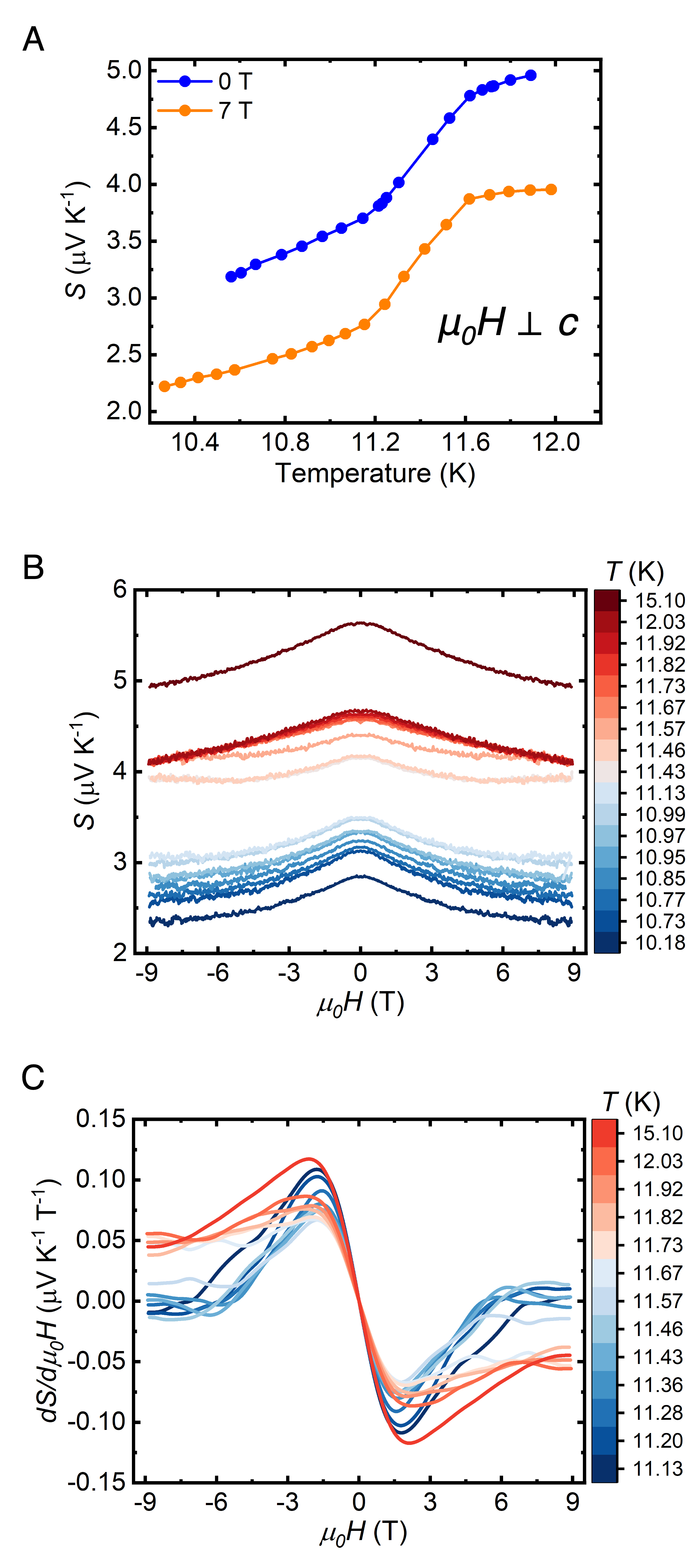}
  \caption{Seebeck ($S$) of \ch{Sm3ZrBi5}. (A) Temperature-dependent Seebeck coefficient at 0 and 7 T with heat applied along the \emph{c} axis. The magnetic transitions are observed at 11.7 and 10.7 K, in agreement with the heat capacity and resistivity measurements. The N\'eel transitions are more pronounced at 7 T but remain at the same temperatures, indicating a higher degree of spin-scattering from the magnetic transition. (B) Seebeck coefficient as a function of applied field, with field applied perpendicular to the \emph{c} axis. There is a clear gap starting at approximately 11.7 K from the first magnetic transition, which can also be observed in the $\frac{dS}{d\mu_0H}$ (C) plot.}
  \label{Seebeck}
  \centering
\end{figure}

To further understand the transport behavior of \ch{Sm3ZrBi5}, we have conducted thermopower (Seebeck) measurements to understand its thermoelectric behavior. Thermopower grants greater insight into how the density of states (DOS) is changing in \ch{Sm3ZrBi5}, along with further elucidating its magnetic and topological behavior. The temperature and field-dependent Seebeck effect is plotted in Figure \ref{Seebeck}, with heat applied along the \emph{c} axis. In Figure \ref{Seebeck}A, the temperature-dependent Seebeck is shown at 0 and 7 T, with field applied perpendicular to \emph{c}. The two magnetic transitions are observed at approximately 11.7 and 10.7 K at both 0 and 7 T, in agreement with the heat capacity and resistivity measurements. The slope of each curve changes significantly during both transitions, which is attributable to the antiferromagnetic order altering the composition of the Fermi surface and changing the DOS. In addition, the Seebeck at 7 T shows that the transition at 11.7 K is much more pronounced, indicating a higher degree of spin-scattering from the initial magnetic transition. Figure \ref{Seebeck}B shows the field-dependent thermopower measurements, where a clear gap is observed at approximately 11.7 K but not as much at 10.7 K, suggesting that the first transition causes a larger change in the DOS than the second. This is corroborated by the $\frac{dS}{d\mu_0H}$ plot (obtained by numerical differentiation) in Figure \ref{Seebeck}C, highlighting the clear separation at the first magnetic transition. Viewed collectively, the thermopower data shows the effect of the magnetic transitions on the electronic structure of \ch{Sm3ZrBi5}, as well as their invariance to applied field for the temperature of each N\'eel transition.

\section{Conclusion}

In conclusion, we have identified hypervalent 1D Pn$^{2-}$ chains as a structural motif of interest for topological materials. In particular, the Ln$_3$MPn$_5$ (Ln = La--Nd; M = Ti, Zr, Hf, Mg, Mn, Nb; Pn = Bi, Sb) family, where Pn acts as both a 1D chain and a ligand anion, has rich chemical diversity, allowing for the study of complex magnetism and topology. \ch{Sm3ZrBi5} was investigated as a newly identified compound from this family. It is a promising material candidate to study the interaction of magnetism and topology due to its complex electronic and magnetic structures. The band structure of \ch{Sm3ZrBi5} has several topologically nontrivial areas in the vicinity of the Fermi level. Because the band structure around the Fermi energy is mostly comprised of Bi $p$ orbitals, SOC has a large effect. Some SOC-induced gaps in the band structure suggest that future spin-resolved experiments will be of interest. Magnetic measurements of the compound could be explained by a quasi-1D magnetic structure with two antiferromagnetic transitions at 11.7 and 10.7 K. These antiferromagnetic transitions are invariant to applied magnetic field up to 9 T, which may be attributable to a magnetic frustration between Sm atoms. These magnetic data are supported by heat capacity and transport measurements, and scattering behavior of \ch{Sm3ZrBi5} is similar to other known topological materials. The identification of Ln$_3$MPn$_5$ as a family of interest for magnetic topological materials will allow for follow-up studies on structural analogues to further understand their electronic, magnetic, and thermal behavior.

\begin{acknowledgement}
This work was supported by the Arnold and Mabel Beckman foundation through a Beckman Young Investigator grant and an AOB postdoctoral fellowship awarded to L.M.S. and J.F.K., respectively. We further acknowledge support by the Packard foundation, the Sloan foundation, and the Gordon and Betty Moore Foundation’s EPIQS initiative through Grant GBMF9064. N.P.O and L.M.S. are supported by a MRSEC award from NSF Grant DMR 2011750. The authors acknowledge the use of Princeton’s
Imaging and Analysis Center, which is partially supported by the
Princeton Center for Complex Materials, a National Science
Foundation (NSF)-MRSEC program (DMR-2011750). J.F.K. would like to thank Dr. Tyler J. Slade, Dr. Daniel G. Chica, and Dr. Rebecca L. Dally for helpful discussions.
\end{acknowledgement}

\begin{suppinfo}
The Supporting Information is available free of charge:

Additional experimental information, crystallographic information, SEM/EDS data, XPS data, projected band structure calculations, additional magnetic data, dilution refrigerator resistance measurements.

\end{suppinfo}

\providecommand{\latin}[1]{#1}
\makeatletter
\providecommand{\doi}
  {\begingroup\let\do\@makeother\dospecials
  \catcode`\{=1 \catcode`\}=2 \doi@aux}
\providecommand{\doi@aux}[1]{\endgroup\texttt{#1}}
\makeatother
\providecommand*\mcitethebibliography{\thebibliography}
\csname @ifundefined\endcsname{endmcitethebibliography}
  {\let\endmcitethebibliography\endthebibliography}{}


For Table of Contents Only
\begin{figure}[H]
\centering
 \includegraphics[width=\textwidth]{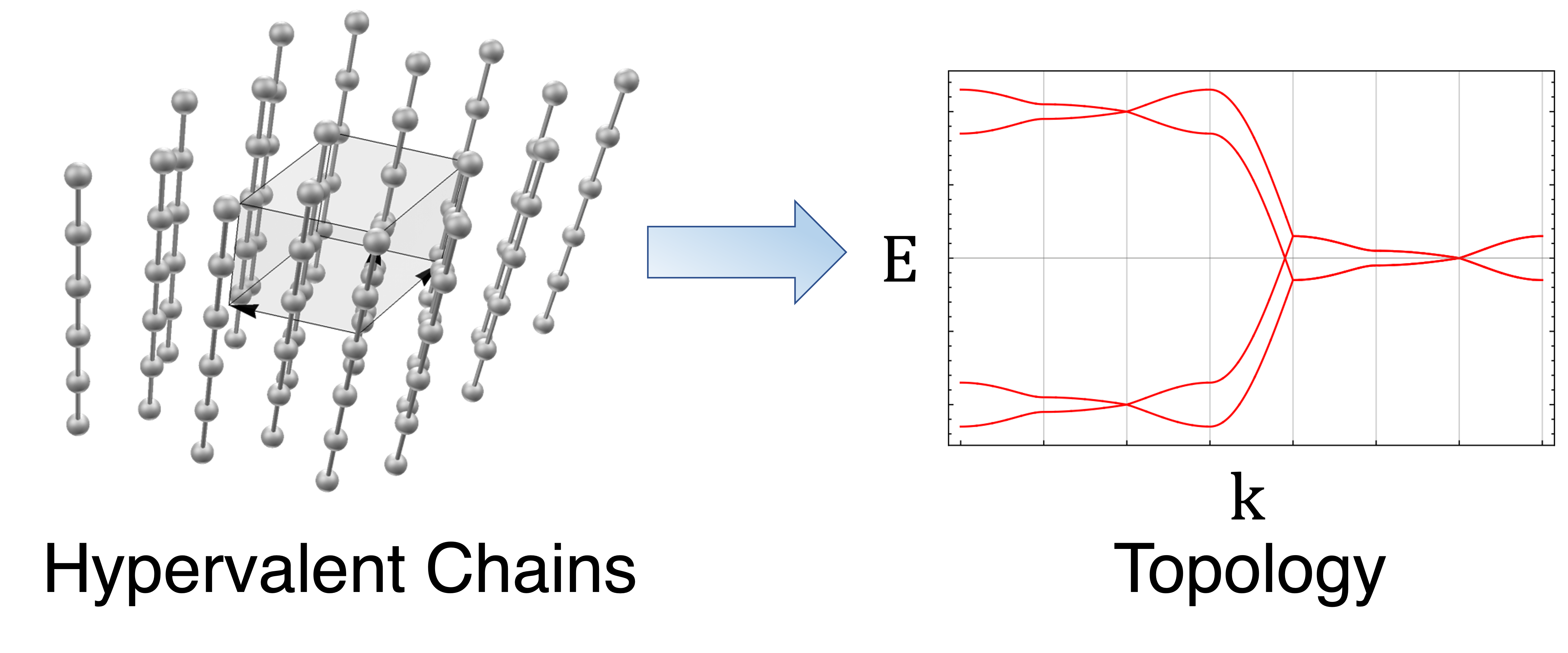}
\end{figure}

\end{document}


\newpage
\tableofcontents
\newpage

\section{Additional Experimental Information}
\subsection{Scanning electron microscopy (SEM) and Energy Dispersive Spectroscopy (EDS)} Scanning electron microscopy and energy dispersive x-ray analysis were performed using a Quanta 200 FEI Environmental-SEM operating at a 20 kV voltage. EDS data were conducted on flux grown crystals of \ch{Sm3ZrBi5} using flat, well-defined faces.

\subsection{X-ray Photoelectron Spectroscopy (XPS)} X-ray photoelectron spectroscopy data were conducted using a ThermoFisher K-$\alpha$ X-Ray Photoelectron Spectrometer. The peaks were calibrated by a C 1$s$ spectrum, where the measured peak at 284.4 eV was moved to the reference value of 284.6 eV. The surface of the \ch{Sm3ZrBi5} crystal was etched by an argon ion gun for 20 seconds to remove any residual Bi flux or surface contamination.

\section{Single Crystal X-ray Diffraction Information}

\begin{table}[H]
  \caption{Crystal data and structure refinement for \ch{Sm3ZrBi5}}
  \label{SCXRD}
  \begin{tabular}{c c}
    \hline
    Empirical formula & \ch{Sm3ZrBi5} \\
    \hline
    Crystal system & Hexagonal  \\
    Space group & \textit{P}$6_3$/\textit{mcm} \\
    Formula weight & 1587.17 \\
    Density (calculated) & 10.426   g/cm$^3$ \\
    a (\si{\angstrom}) & 9.5386(3)  \\
    b (\si{\angstrom}) & 9.5386(3)  \\
    c (\si{\angstrom}) & 6.4158(3)  \\
    $\alpha$ ($^\circ$) & 90 \\
    $\beta$ ($^\circ$) & 90 \\
    $\gamma$ ($^\circ$) & 120 \\
    Volume ($\si{\angstrom}^3$) & 505.53(4)  \\
    Z & 2 \\
    Temperature (K) & 298(1) \\
    F(000) & 1282 \\
    Reflections collected & 5836 \\
    $\Theta$ ($^\circ$) & 2.466 - 24.885 \\
    Crystal size (mm$^3$) & 0.108 $\times$ 0.069 $\times$ 0.047 \\
    Absorption coefficient (mm$^{-1}$) & 104.71  \\
    R$_{int}$ & 0.0692 \\
    Refinement method & Full-matrix least-squares on F$^2$ \\
    Final R indices (R$_{obs}$/wR$_{obs}$) & 0.0299/0.0648 \\
    R indices (all data) (R$_{all}$/wR$_{all}$) & 0.0301/0.0650 \\
    Goodness-of-fit & 1.279 \\
    Extinction coefficient & 0.0030(4) \\
    Largest diff. peak and hole (e$\cdot\si{\angstrom}^{-3}$) & 2.581 and -2.767  \\
    \hline
    $R=\frac{\Sigma||F_{o}|-|F_{C}||}{\Sigma|F_{o}|}$ & \\
    $wR=[\frac{\Sigma[w(|F_{o}|^{2}-|F_{c}|^{2})^{2}]}{\Sigma[w(|F_{o}|^{4})]}]^{1/2}$ &
    $w=\frac{1}{\sigma^2(F_{o}^{2})+(0.0298P)^{2}+10.4309P}$ $P=\frac{F_{o}^{2}+2F_{c}^{2}}{3}$\\
 \hline
  \end{tabular}
\end{table}

\begin{table}[H]
  \caption{Atomic coordinates (x10$^4$) and equivalent isotropic displacement parameters ({\AA}$^2$x10$^3$) for Sm$_3$ZrBi$_5$ at room temperature with estimated standard deviations in parentheses}
  \label{Coord}
  \begin{tabular}{c c c c c c}
    \hline
    Label & x & y & z & Occupancy & U$_{eq}$* \\
    \hline
    Bi(01) & -2677(1) & 0 & 2500 & 1 & 29(1) \\
    Bi(02) & 3333 & 6667 & 5000 & 1 & 30(1) \\
    Sm(03) & 0 & 3818(2) & 2500 & 1 & 31(1) \\
    Zr(04) & 0 & 0 & 5000 & 1 & 28(1) \\
    \hline
  \end{tabular}\\
  *U$_{eq}$ is defined as one third of the trace of the orthogonalized U$_{ij}$ tensor.
\end{table}

\begin{table}[H]
  \caption{Anisotropic displacement parameters ({\AA}$^2$x10$^3$) for Sm$_3$ZrBi$_5$ at room temperature with estimated standard deviations in parentheses}
  \label{ADP}
  \begin{tabular}{c c c c c c c}
    \hline
    Label & U$_{11}$ & U$_{22}$ & U$_{33}$ & U$_{12}$ & U$_{13}$ & U$_{23}$\\
    \hline
    Bi(01) & 28(1) & 30(1) & 29(1) & 15(1) & 0 & 0\\
    Bi(02) & 30(1) & 30(1) & 28(1) & 15(1) & 0 & 0  \\
    Sm(03) & 33(1) & 30(1) & 31(1) & 16(1) & 0 & 0 \\
    Zr(04) & 28(2) & 28(2) & 28(2) & 14(1) & 0 & 0 \\
    \hline
  \end{tabular}\\
  The anisotropic displacement factor exponent takes the form:
  
 --2$\pi^2$[h$^2$a$^{*2}$U$_{11}$ + ... + 2hka$^*$b$^*$U$_{12}$].
\end{table}

\section{Additional Experimental Data}

\begin{figure}[hbt!]
  \includegraphics[width=\textwidth]{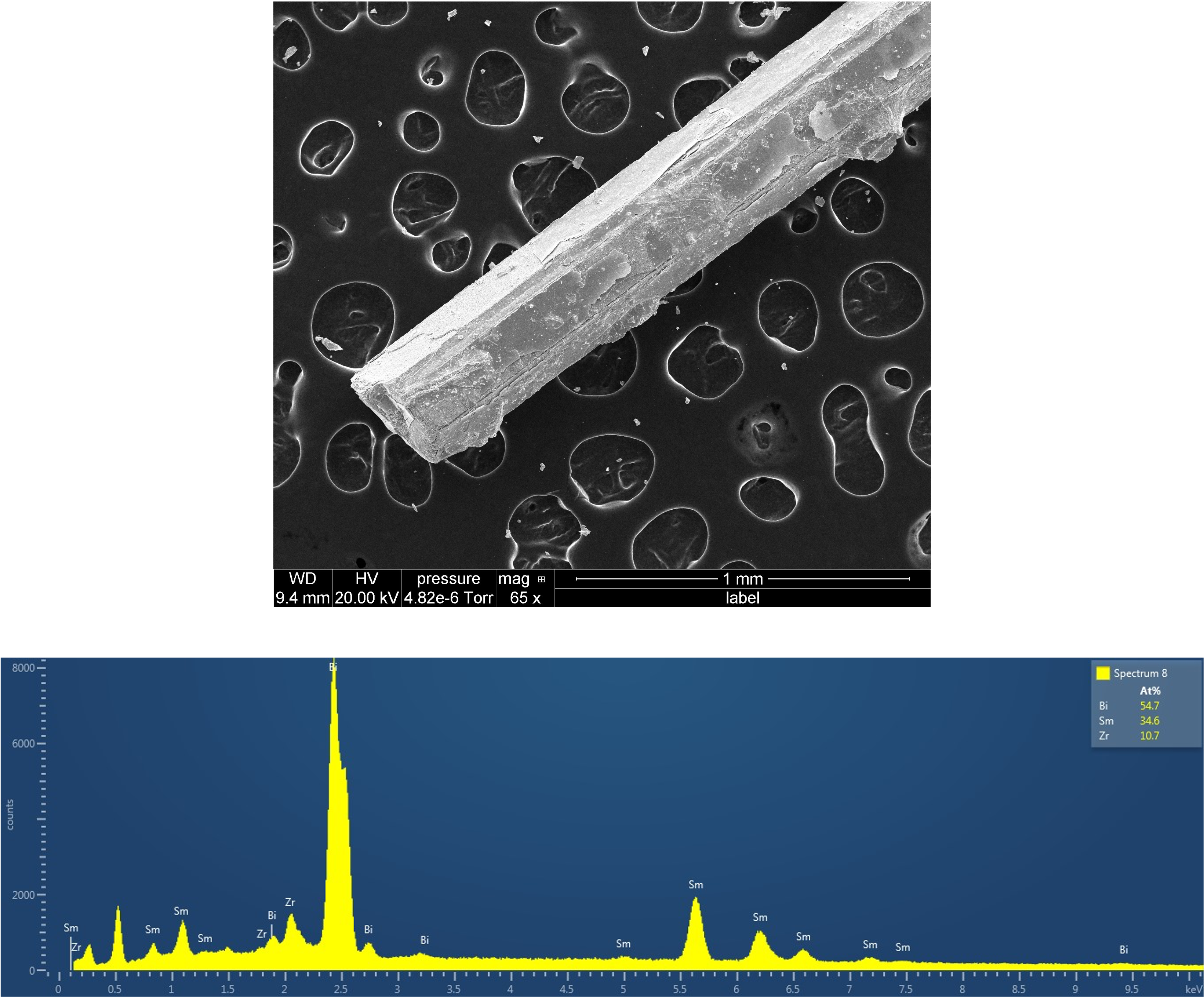}
  \caption{SEM and EDS of \ch{Sm3ZrBi5}}
  \label{SEM_EDS}
  \centering
\end{figure}

\begin{figure}[hbt!]
  \includegraphics[width=0.5\textwidth]{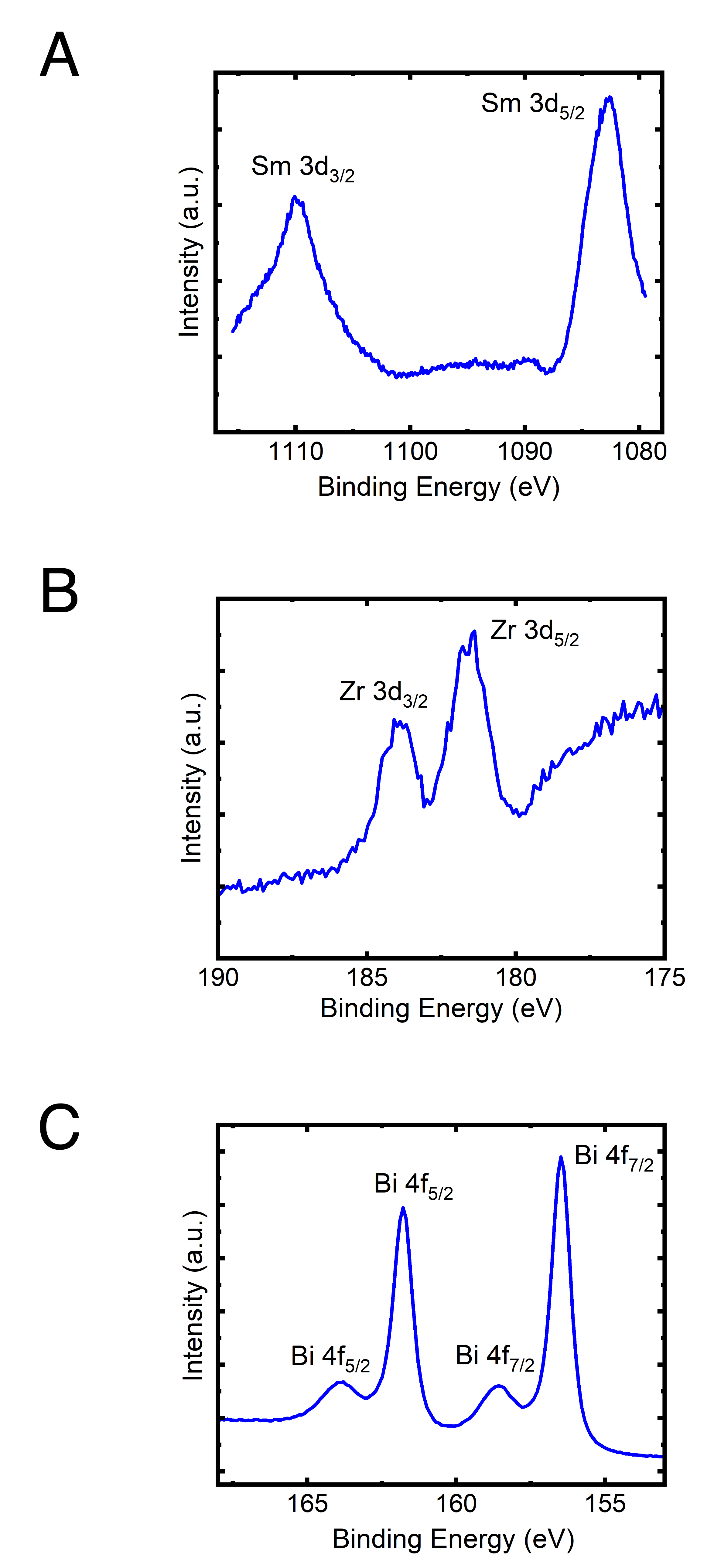}
  \caption{XPS of \ch{Sm3ZrBi5}}
  \label{XPS}
  \centering
\end{figure}

\begin{figure}[hbt!]
  \includegraphics[width=\textwidth]{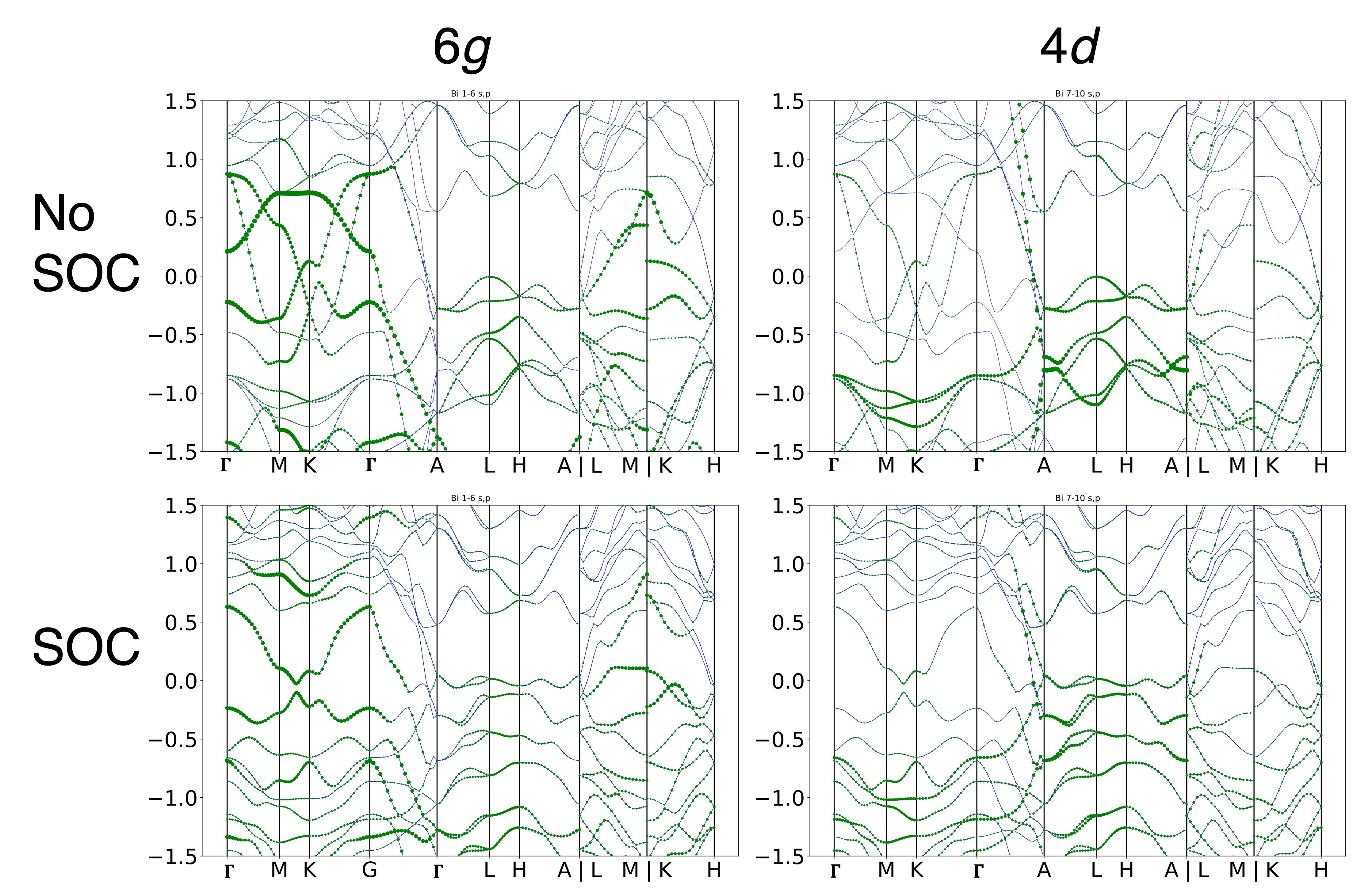}
  \caption{Orbital projected (\emph{e.g.} ``fat band'') band structure for \ch{Sm3ZrBi5} with and without SOC at the 4$d$ and 6$g$ Wyckoff positions of Bi. The projected band structure is composed of Bi s and p orbital contributions. The rows indicate without and with SOC, and the columns indicate the specific Wyckoff site for each band structure.}
  \label{Fatbandsp}
  \centering
\end{figure}

\begin{figure}[hbt!]
  \includegraphics[width=\textwidth]{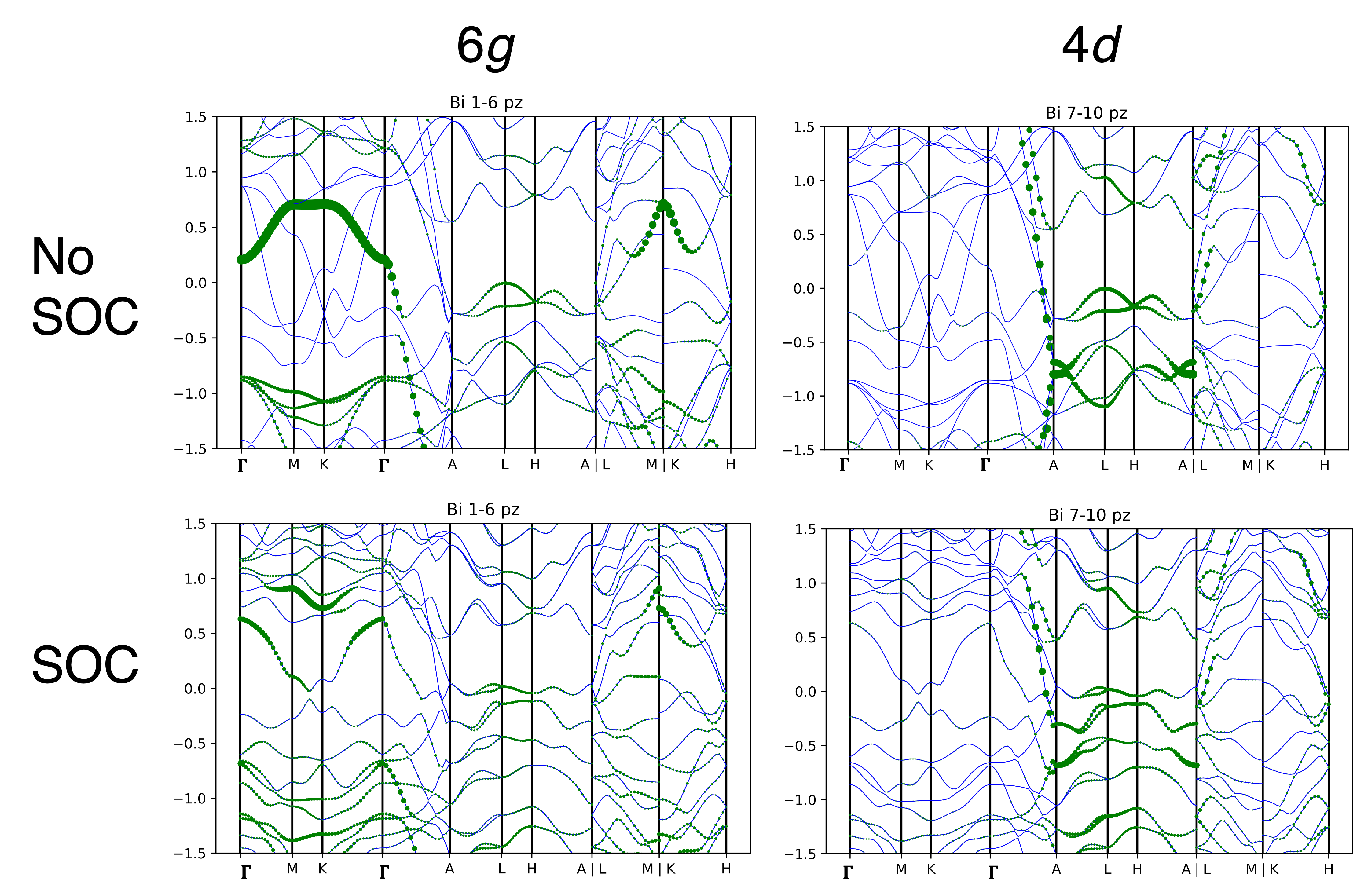}
  \caption{Orbital projected (\emph{e.g.} ``fat band'') band structure for \ch{Sm3ZrBi5} with and without SOC at the 4$d$ and 6$g$ Wyckoff positions of Bi. The projected band structure is composed of Bi $p_z$ orbital contributions. The rows indicate without and with SOC, and the columns indicate the specific Wyckoff site for each band structure.}
  \label{Fatbandpz}
  \centering
\end{figure}

\begin{figure}[hbt!]
  \includegraphics[width=\textwidth]{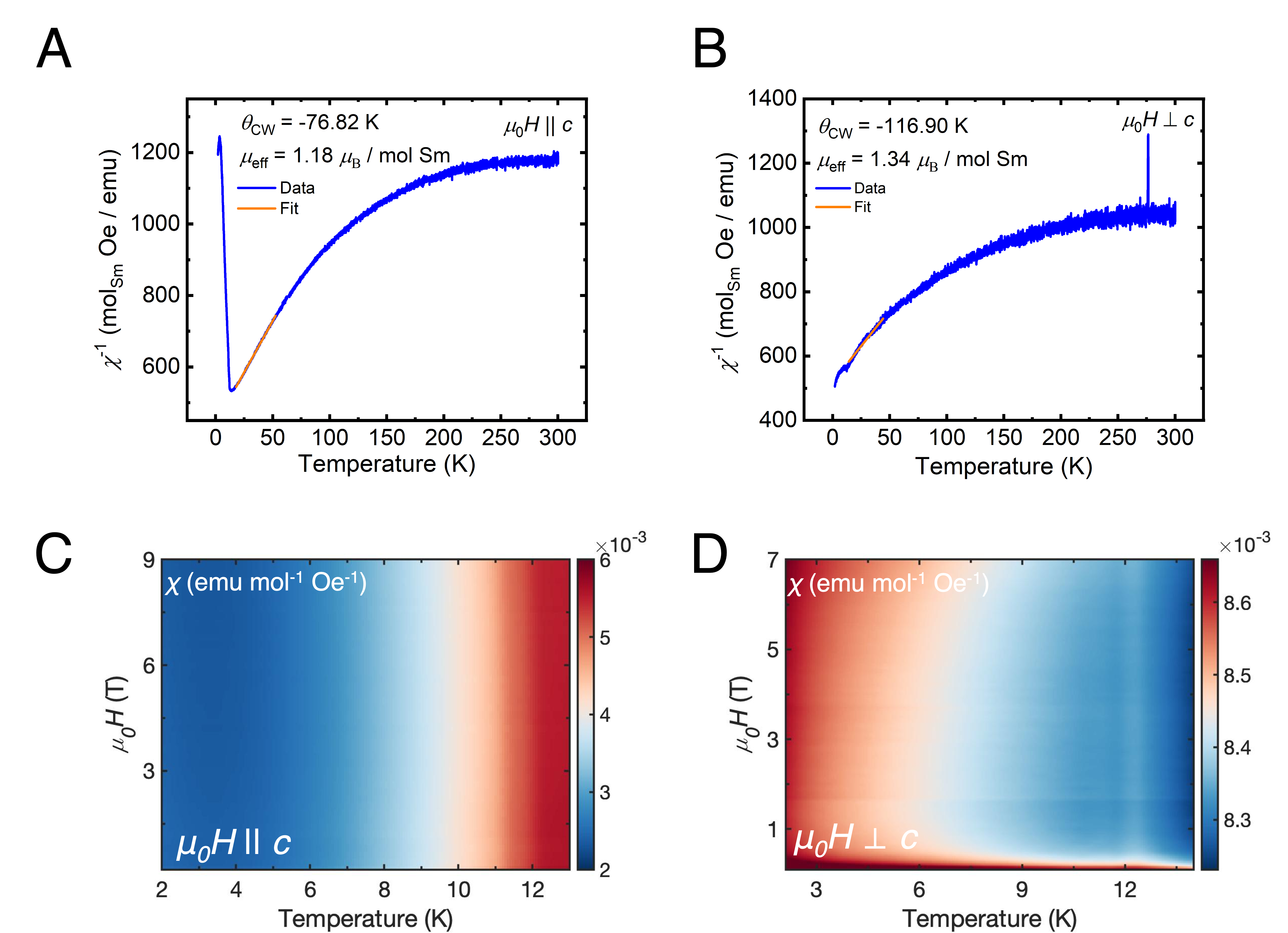}
  \caption{Additional magnetic data for \ch{Sm3ZrBi5}. (A--B) Curie-Weiss fits with field applied parallel (A) and perpendicular (B) to the \emph{c} direction. (C--D) Magnetic phase diagrams for magnetic susceptibility (\emph{$\chi$}) with field applied parallel (C) and perpendicular (D) to the \emph{c} direction.}
  \label{Magnetism_SI}
  \centering
\end{figure}

\begin{figure}[hbt!]
  \includegraphics[width=\textwidth]{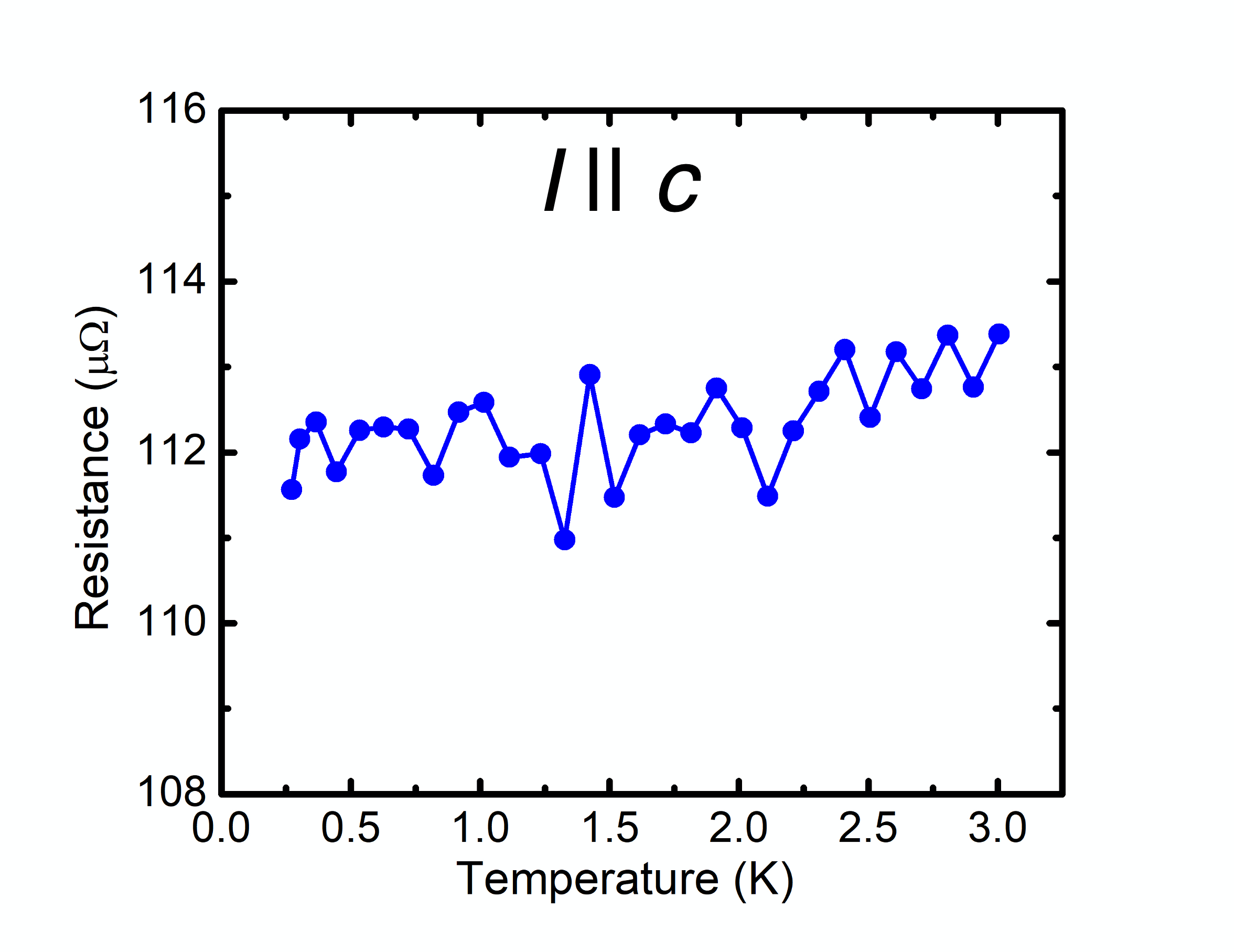}
  \caption{Dilution refrigerator resistance data for \ch{Sm3ZrBi5} (with current applied along the \emph{c} axis) down to approximately 0.3 K, showing no additional magnetic or superconducting transitions.}
  \label{DilutionFridge}
  \centering
\end{figure}
